\documentclass[12pt,preprint]{aastex}

\usepackage{graphicx}

\newcommand{\chandra}{{\it Chandra}}

\newcommand{\xmm}{{\it XMM-Newton}}
\newcommand{\HST}{{\it Hubble Space Telescope}}
\newcommand{\hst}{{\it HST}}

\newcommand{\ha}{H$\alpha$}
\newcommand{\nii}{\ion{N}{2}}

\newcommand{\sii}{\ion{S}{2}}
\newcommand{\oiii}{\ion{O}{3}}
\newcommand{\oii}{\ion{O}{2}}
\newcommand{\oi}{\ion{O}{1}}

\shorttitle{Multiwavelength Analysis of the Physical Structure of N49}
\shortauthors{Bilikova et al.}

\begin{document}


\title{Supernova Remnants in the Large Magellanic Clouds IX: \\
    Multiwavelength Analysis of the Physical Structure of N49}


\author{J. Bilikova, R. N. M. Williams, Y.-H. Chu, R. A. Gruendl and
B. F. Lundgren}
\affil{Astronomy Department, University of Illinois at Urbana-Champaign \\
 1002 W. Green St. Urbana, IL 61801}



\begin{abstract}

We present a multiwavelength analysis of the supernova remnant N49 in the Large
Magellanic Cloud. Using high-resolution {\it Hubble Space Telescope} WFPC2 
images of H$\alpha$, [\sii] and [\oiii] emission, we study the morphology of 
the remnant and calculate the rms electron densities in different regions. 
We detect an offset of [\oiii] and H$\alpha$ emission peaks of about 0\farcs5, 
and discuss possible scenarios that could give rise to such high values.
The kinematics of the remnant is analyzed by matching individual kinematic
features in the echelle spectra obtained at CTIO with the morphological
features revealed in the WFPC2 images.  We detect narrow H$\alpha$ emission
component and identify it as the diffuse pre-shock recombination radiation,
and discrete broad emission features that correspond to the shocked gas in
filaments.  The overall expansion of the remnant is about 250 km s$^{-1}$.
The dense clouds are shocked up to line-of-sight velocities of 250 km s$^{-1}$ 
and the less dense gas up to 300 km s$^{-1}$.  A few cloudlets have even
higher radial velocities, reaching up to 350 km s$^{-1}$.  We confirm
the presence of the cavity in the remnant, and identify the center of 
explosion. 
Using archival  {\it Chandra} and {\it XMM-Newton} data, we observe the same 
trends in surface brightness distribution for the optical and X-ray images.
We carry out the spectral analysis of three regions that represent the most 
significant optical features.

\end{abstract}



\section{Introduction}

N49 is the most optically bright supernova remnant (SNR) in the Large Magellanic Cloud (LMC).  In contrast to a classic shell-type SNR, N49 exhibits largely asymmetric morphology at optical wavelengths. The increased brightness of this SNR in the southeast and its unique filamentary structure at optical wavelengths has long set N49 apart from other well-understood SNRs.  Inhomogeneities in the optical morphology  and kinematics of N49 have been observed in the past \citep{Shull83,CK88,V92}.  However, the fine structure resolved from these kinematic observations was difficult to reconcile with the optical images taken from the ground.   Recent mapping of molecular clouds in the region suggests that this SNR is expanding into a denser region toward the southeast, which has caused the asymmetric brightness enhancements in the shell \citep{B97}. 

The presence of such a filamentary structure suggests that dense material surrounded the progenitor star at the time of its supernova.  In the case of a massive progenitor star, its strong stellar wind would have cleared most material from the environment during its lifetime.   Considering the evidence for a relatively dense environment surrounding N49, a B-type star, which would lack a strong stellar wind, is the most likely candidate for the N49 progenitor \citep{Shull83}.   

This SNR also appears notably bright at the X-ray \citep{RW99, Park03} and radio \citep{D95} wavelengths, in which the complete spherical shell of the remnant is clearly visible.  While the overall structure appears symmetric at X-ray wavelengths, the brightness peaks in the southeastern region as in the optical.  In the {\it Chandra} observations, most of the optically bright filamentary structures have only diffuse X-ray counterparts.  These observations also revealed a hard X-ray knot extending beyond the main shell in the southwest of N49, which has been suggested to be a fragment of the explosion accelerated beyond the shock boundary. 

An additional point of interest is the existence of a known soft gamma-ray repeater, SGR 0526-66, within the boundary of N49.  Recent {\it Chandra} images have resolved a point source, suggested to be the X-ray counterpart to this SGR \citep{K03}.  No optical or radio counterparts of this SGR have been detected \citep{D95,Setal98,Fetal98}, and the SGR does not seem to have strong influence on the overall optical morphology of the remnant.

In order to understand the kinematics of the remnant and the distribution of X-ray emission in terms of the detailed physical structure of the N49 SNR, we have obtained high-resolution {\HST} ({\it HST}) images in the H$\alpha$, [\oiii], and [\sii] emission lines, which reveal N49's bright filamentary structures with high precision. We have supplemented existing high-dispersion spectroscopy \citep{CK88} with additional deep spectra for N49 to investigate the full extent of velocity distribution.  In this paper, we analyze the above observations in conjunction with archival {\it Chandra} and {\it XMM-Newton} observations for a comprehensive multi-wavelength study of the physical structure of N49.  \S 2 describes the included observations, \S 3 analyzes the {\it HST} emission line images and the implications of the surface brightness and the line ratios, \S 4 analyzes the nebular kinematics and correlates the kinematic and morphological features, and \S 5 compares the X-ray and optical morphologies and discusses the relative distribution of hot and cool gas, and a summary is given in \S 6.

\section{Observations}


\subsection{Optical Emission-Line Images}

We imaged N49 using the {\hst}  Wide Field Planetary
Camera 2 (WFPC2) with three filters: F656N 
(H$\alpha$), F673N ([\ion{S}{2}] $\lambda\lambda$6717, 6731) and F502N
([\ion{O}{3}] $\lambda$5007).  The dates and exposure times of these observations are presented in Table 1. The data were reduced using the  \textsc{stsdas} package within \textsc{iraf}\footnote{Image Reduction
Analysis Facility, maintained by NOAO}. 
The continuum was not subtracted, as the line emission was so dominant
that the presence of continuum would not affect our conclusions.
Cosmic-ray events were removed by
combining multiple exposures, and the resulting files were bias-subtracted.  
Each image was divided by its exposure time to produce a count-rate map.
These maps in turn were multiplied by a conversion factor, given by the
\textsc{photflam} parameter  in the image header, to produce flux-density 
maps. We used the \textsc{synphot} task to determine widths for each filter, 
and multiplied the flux-density maps by the corresponding filter widths to produce flux 
maps. The resulting files were then mosaicked together for the final images.

\subsection{Optical Spectroscopy}

We obtained high-dispersion spectra of N49 with the echelle
spectrograph on the Blanco 4m telescope at Cerro Tololo Inter-American 
Observatory (CTIO) in three observing runs.
A journal of our observations is given in Table 2.Ê We used the 
79 line mm$^{-1}$ echelle grating in the single-order, long-slit 
observing configuration, with a flat mirror replacing the cross 
disperser and a post-slit \ha\ filter ($\lambda_c$ = 6563 \AA, 
$\Delta \lambda$ = 75 \AA) inserted to isolate a
single order.

Due to the vignetting limitations imposed by the optics, the useful 
spatial coverage is $\sim$3$'$.
The spectral coverage includes both the \ha\ line and the neighboring
[\nii]~$\lambda\lambda$6548, 6583 lines. We used a slit width of 250
$\mu$m (1\farcs64) for all observations. 
The spectral dispersion and wavelength were calibrated by a Th-Ar lamp 
exposure taken in the beginning of each night, and fine-tuned using the 
geocoronal \ha\ and telluric OH lines present in the echellograms.

The 1986 observing run used the Air Schmidt camera and a 
GEC 385$\times586$ CCD.  The 22 $\mu$m pixel size corresponds 
to about 0.21 \AA\ and 0\farcs635 along the dispersion and spatial axes, 
respectively.
The instrumental profile had a FWHM of $21\pm0.5$ km s$^{-1}$, 
as measured from the Th-Ar calibration and sky lines.
The echelle spectra obtained in this observing run have been reported by
 \citet{CK88}.

The 1995 observing run used the red long-focus camera and a Tek 
2048$\times$2048 CCD. 
The 24 $\mu$m pixel size corresponds to roughly 0.082 \AA\ and
0\farcs26 along the dispersion and spatial axes, respectively.
The instrumental profile had a FWHM of $16.1\pm0.8$ km s$^{-1}$.

The observations made in 2000 were recorded using the red  long-focus 
camera and the SITe2K\#6 2048$\times$2048 CCD.
The 24 $\mu$m pixel size corresponds to roughly 0.082 \AA\ and
 0\farcs26 along the diespersion and spatial axes, respectively. 
The instrumental profile had a FWHM of $13.5\pm0.5$ km s$^{-1}$.

\subsection{X-ray Images and Spectroscopy}

X-ray observations of N49 were made with both the \xmm\ and \chandra\
observatories.  We retrieved the pipeline-processed \xmm\ data from the
Science Operations Centre (SOC). Observations were made simultaneously
with multiple \xmm\ instruments;  in this paper we will concentrate on
the European Photon Imaging Camera (EPIC) MOS and pn detectors. The
data were taken in 2001 April (Observation ID 0113000201; 26.9 ks; 
PI: Kaastra) over multiple time intervals, using the Thick filter for two 
EPIC-MOS1 exposures and the Medium filter for the remaining EPIC-MOS 
and EPIC-pn exposures. 
Initial reduction and analysis were carried out using the
Science Analysis Software package provided by the SOC, with
subsequent spectral analysis in \textsc{xspec}.

The data were filtered to remove high background times or poor event
grades and merged to produce a single dataset for each instrument with
effective exposure times of 14.3 ks for EPIC-MOS1,
19.5 ks for EPIC-MOS2 and 0.8 ks for EPIC-pn.  Images were
then extracted from the filtered merged event files.  
Spectra were extracted separately from each exposure, to avoid introducing 
distortions due to the different filters used.  
Background regions immediately surrounding the SNR but free of point 
sources were used to produce background spectra, which were then scaled 
and subtracted from the source spectra. Spectra from different 
exposures were jointly fitted.

\chandra\ observations used the Advanced CCD Imaging
Spectrometer (ACIS), primarily the S3 back-illuminated chip. Datasets
were sequence number 500107, observation 1041 (34.3 ks; PI: Garmire) and sequence
number 500134, observation 2515 (7.0 ks; PI: Kulkarni).  Data were reprocessed
following procedures recommended by the Chandra X-ray Center:
we removed afterglow corrections, generated new bad pixel files, and
applied corrections for charge-transfer inefficiency and
time-dependent gain for an instrument temperature of 120 K. The two
datasets were filtered for high-background times and poor event
grades, resulting in total ``good time" intervals of 33.9 and 6.9 ks.

Spectral results were extracted from the new event files. Background regions
were taken from source-free areas surrounding the SNR, and the spectra
of these background regions were scaled and subtracted from the source
spectra.  Individual spectra for regions of interest, selected by comparison
of optical and X-ray images, and the corresponding primary and auxiliary 
response files were extracted with the \textsc{ciao} acisspec script and 
analyzed in \textsc{xspec}. 
Spectra were rebinned by spectral energy to achieve a signal-to-noise ratio 
of 4 in each bin.

\section{Analysis of the HST Emission Line Images}
\subsection{H$\alpha$ Surface Brightnesses and Electron Densities}

\begin{figure}
\begin{center}
\includegraphics[width=1.0\textwidth]{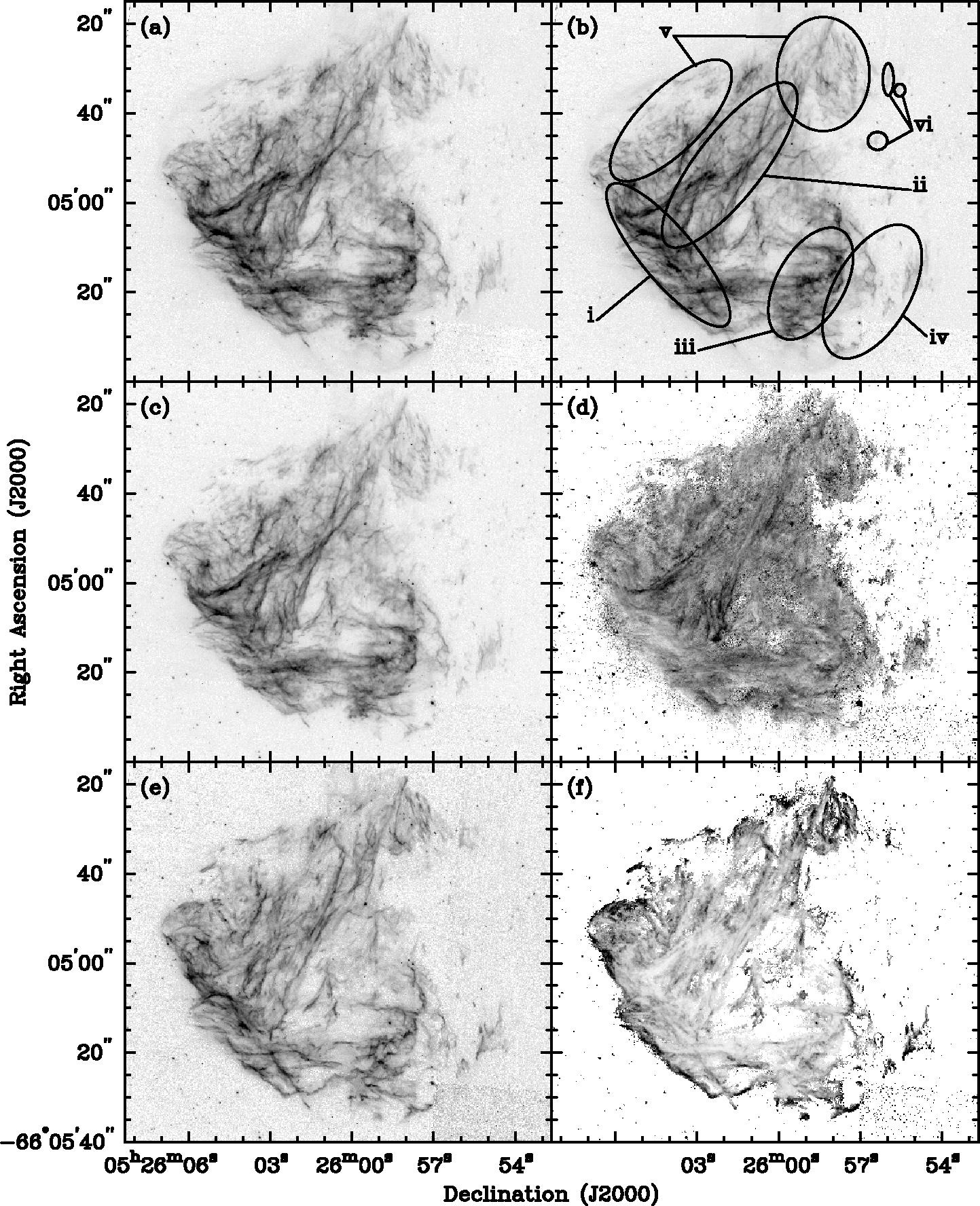}
\caption{Mosaicked and flux-calibrated {\it HST} WFPC2 images and ratio maps 
of N49 displayed in linear greyscale.The panels are (a) H$\alpha$ image, 
(b) morphological features marked on the H$\alpha$ image, 
(c) [\sii] emission line image, (d) [\sii]/H$\alpha$
ratio map, (e) [\oiii] emission line image, (f) [\oiii]/H$\alpha$ ratio map.} 
\label{fig:has2o3}
\end{center}
\end{figure}

The {\it HST} H$\alpha$ image (Fig. \ref{fig:has2o3}a) displays a complex 
network of filaments with the brightest emission located toward the southeast.
In Figure \ref{fig:has2o3}b, we have marked the morphological features of the
remnant, which will henceforward be referred to as features i to vi.
The most prominent feature is the SE Ridge (feature i), a ridge of bright 
filaments in the southeast quadrant, tracing a line from east to south 
of the SNR. 
From the SE Ridge, straight bright filaments extend toward the northwest over 
lengths of 15--30$''$ (ii).
A network of bright filaments is present in the southwestern part of the
remnant (iii), where the
filaments are shorter than those in the SE Ridge.
Beyond this network toward the outer edge of the remnant, there exist a 
number 
of fainter filaments (iv).
Faint filaments are also seen in the northern and northeastern parts of the
SNR (v). No bright filaments exist in the western part of the remnant.

The bright filaments are enveloped by a diffuse H$\alpha$
emission component.
This diffuse emission extends beyond the
radio/X-ray \citep{RW99, Park03, D95} boundary along the circumference from 
the north through the east to the south.
On the west side of the remnant, only faint patches of H$\alpha$
emission exist (feature vi).
While the outermost emission patches follow the X-ray/radio
boundary, they
 do not appear to be
embedded in a diffuse emission component, which is either absent
 or below the detection limit.

Due to its presence beyond the shock boundary, we identify the diffuse 
H$\alpha$ emission as the recombination radiation of hydrogen ionized 
by the UV precursor in the pre-shock region.

The H$\alpha$ surface brightness ($SB$) of a 10$^4$ K ionized
gas can be expressed as
\begin{equation}
SB = 1.9 \times 10^{-18} n_{\rm e}^2~ L_{\rm pc} ~~
{\rm ergs ~s^{-1} ~cm^{-2} ~arcsec^{-2}},
\end{equation}
where $n_{\rm e}$ is the electron density in cm$^{-3}$ and
 $L_{\rm pc}$ is the emitting path length in pc.
Using the flux-calibrated {\it HST} H$\alpha$ image\footnote{The WFPC2 
H$\alpha$ filter (F656N) also transmits the [\nii] $\lambda$6548 line, 
which is $\sim$8\% as bright as the H$\alpha$ 
line in N49 \citep{Vancura92M}. This [\nii] contamination has been corrected 
for.} 
and reasonable estimates
of the emitting path lengths, we may determine the gas densities.
For a filament, we adopt its length as upper limit and its width as lower
limit for the emitting path length.
For a diffuse emission region, we adopt its average diameter as the emitting
path length.

In the brightest filaments, such as the SE Ridge,
the long
filaments extending to the northwest, and the bright
short filaments in the southwest,
$SB$ is $\sim (1 \pm 0.4) \times 10^{-13}$
ergs s$^{-1}$cm$^{-2}$ arcsec$^{-2}$, with the peak emission
reaching
$2 \times 10^{-13}$ ergs s$^{-1}$cm$^{-2}$arcsec$^{-2}$.
The rms densities in these bright filaments are between the limits of
265 $\pm$ 20 and 515 $\pm$ 20 cm$^{-3}$, adopting the length and width of the 
filaments as the emitting path lengths, respectively.

The filaments in the central and northeastern regions, as well as the 
brighter northern filaments, are slightly fainter than the SE Ridge, 
with $SB$ of
$(3-8) \times 10^{-14}$ ergs s$^{-1}$cm$^{-2}$arcsec$^{-2}$.
The rms electron densities in these filaments are between
the limits of 115 $\pm$ 20 and 310 $\pm$ 20 cm$^{-3}$.
The $SB$ of the fainter filaments in the remnant, such as the northernmost
filaments, is $(0.9 - 2) \times 10^{-14}$
 ergs s$^{-1}$cm$^{-2}$arcsec$^{-2}$.
The rms densities here are between the limits of 70 $\pm$ 20 and 225 $\pm$ 
25 cm$^{-3}$.

The faint patches in the western quadrant have $SB$
of $(1-4) \times 10^{-15}$ ergs s$^{-1}$cm$^{-2}$arcsec$^{-2}$.
Adopting their average diameters as the emitting path lengths, the rms 
densities 
in the patches are 90 $\pm$ 10 cm$^{-3}$.
The diffuse H$\alpha$ emission that envelops the bright filaments has a
similar $SB$, but a larger emitting path length.
Approximating the emitting path length by the radius of the remnant, 
we get rms densities of about 10 $\pm$ 5 cm$^{-3}$.

Our derived rms densities are significantly lower than the densities of
1000 -- 1800 cm$^{-3}$ calculated using
the [\sii] doublet ratios reported by \citet{Vancura92M}.
This difference can be explained by the different weightings
placed on density by these two methods.
The flux of an emission line is proportional to $\int n_{\rm e}^2~dL$,
thus an over-dense region along a line of sight will dominate
the emission and dictate the electron density determined from the
[\sii] doublet ratio.
On the other hand, the rms density is derived from the emissivity
averaged over the entire emitting path length, and therefore is
always lower than the peak value determined from the [\sii] doublet.
In general, the density determined from the [\sii] doublet is
representative of the densest regions, while the rms density better 
represents the density over the entire emission path length.
The rms densities are particularly useful for region whose density 
is $<$ 100 cm$^{-3}$, the low-density limit of the [\sii] doublet, such as 
the faint emission regions.


\subsection{[\sii] Emission and [\sii]/H$\alpha$ Ratios }

Overall, the morphology of N49 in the [\sii] line (Fig. \ref{fig:has2o3}c) 
resembles that in the H$\alpha$ line.
For example, the bright SE Ridge, the straight bright filaments extending to 
the northwest, the fainter filaments in the north and west, and the faint
emission patches in the west are all observed in the [\sii] image.
The most notable morphological difference between the [\sii] and H$\alpha$ 
images is the lack of a diffuse [\sii] emission component enveloping 
the bright filaments.

As the morphologies of N49 in the H$\alpha$ and [\sii] lines are similar, 
the ratio map (Fig. \ref{fig:has2o3}d) is fairly uniform, with a mean value 
of 0.75  $\pm$ 0.05.
However, there are variations in this ratio, giving an overall range of 
0.4 -- 1.4. These variations result from different shock conditions.
The highest ratios are found in the bright filaments extending northward from
the SE Ridge, where the ratios reach values of 1 -- 1.4.
The lowest ratios can be seen in the outermost limb of the remnant, where 
the ratios are below 0.4, which may be due to the lack of a diffuse [\sii] 
component.
In addition, we find low ratios in the SE Ridge, where the H$\alpha$ 
filaments are the brightest.

\subsection{[\oiii] Emission and [\oiii]/H$\alpha$ Ratios}

The [\oiii] emission (Fig. \ref{fig:has2o3}e) is filamentary and peaks in the 
southeast just like the H$\alpha$ and [\sii], but the filamentary
structures differ in details.
The [\oiii] emission associated with bright H$\alpha$ and
[\sii] filaments is usually concentrated in faint thin filaments
and shows an offset toward the leading edge of the shock fronts.
The most extreme difference is exhibited in the  bright H$\alpha$ 
filament extending from the northern tip of the SE Ridge toward 
northwest, where the [\oiii] emission is absent.

These and other morphological differences are best demonstrated
 by the [\oiii]/H$\alpha$ ratio map (Fig. \ref{fig:has2o3}f).
The ratios generally increase from the inner to the outer part of the remnant,
spanning a wide range from nearly 0 to about 2.5.
The ratios in the inner portions of the remnant range from 0 to about 0.5,
with a few regions reaching ratios of 0.7.
The ratio map in the inner portions of the remnant shows a filamentary 
structure that is not as sharp as the direct images.
The brighter and fainter ``filaments'' on the ratio map have values of 
0.4 -- 0.7, and 0.05 -- 0.15, respectively.
Generally, the filaments bright in H$\alpha$ have very low [\oiii]/H$\alpha$
ratios. This can be seen in the SE Ridge, where the brightest
H$\alpha$ filaments are faint on the ratio map.
Ratios above 0.5 are found primarily in the outer portions of the remnant.
The most prominent feature of the ratio map is the increase in the ratios 
toward the outermost edge of the remnant, where
they reach values between 0.8 and 1.5. A few small patches in the north, 
northeast and southwest reach the values of up to 2.5 on their outer edge.

\begin{figure}
\begin{center}
\includegraphics[width=0.9\textwidth]{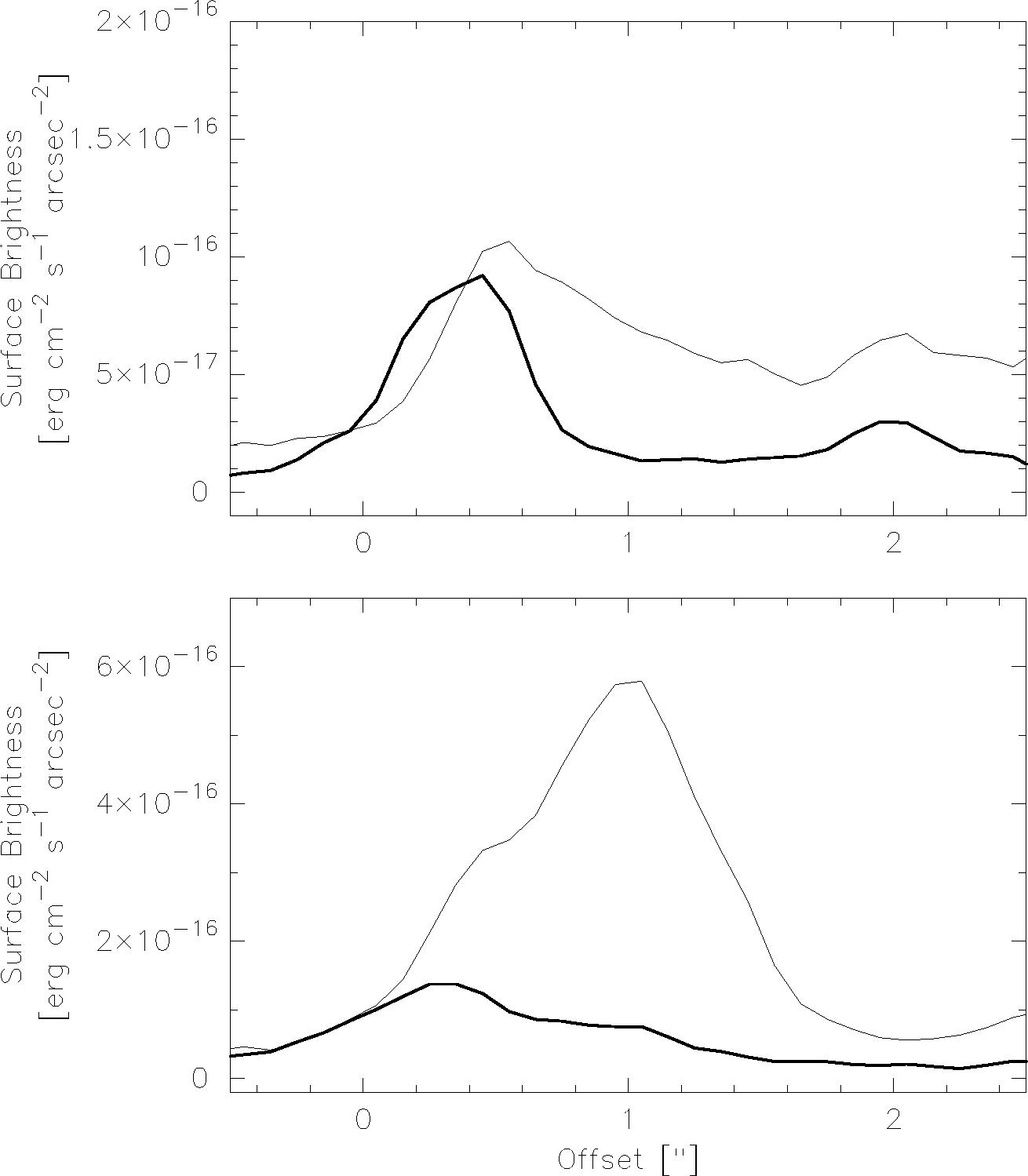}
\caption{Surface brightness profiles of H$\alpha$ (thin line) and [\oiii]
(thick line) emission across selected filaments. 
The coordinates for the positions marked as zero are
05$^{\rm h}$26$^{\rm m}$01$\fs$34, $-$66$^{\circ}$05$'$29$\farcs$1
(J2000), at PA = 30$^{\circ}$ for the top panel, and 
05$^{\rm h}$26$^{\rm m}$04$\fs$4, $-$66$^{\circ}$05$'$18$\farcs$7 
(J2000) at PA = 135$^{\circ}$ for the bottom panel.} 
\label{fig:sbprofile}
\end{center}
\end{figure}

\begin{figure}
\begin{center}
\includegraphics[width=0.49\textwidth]{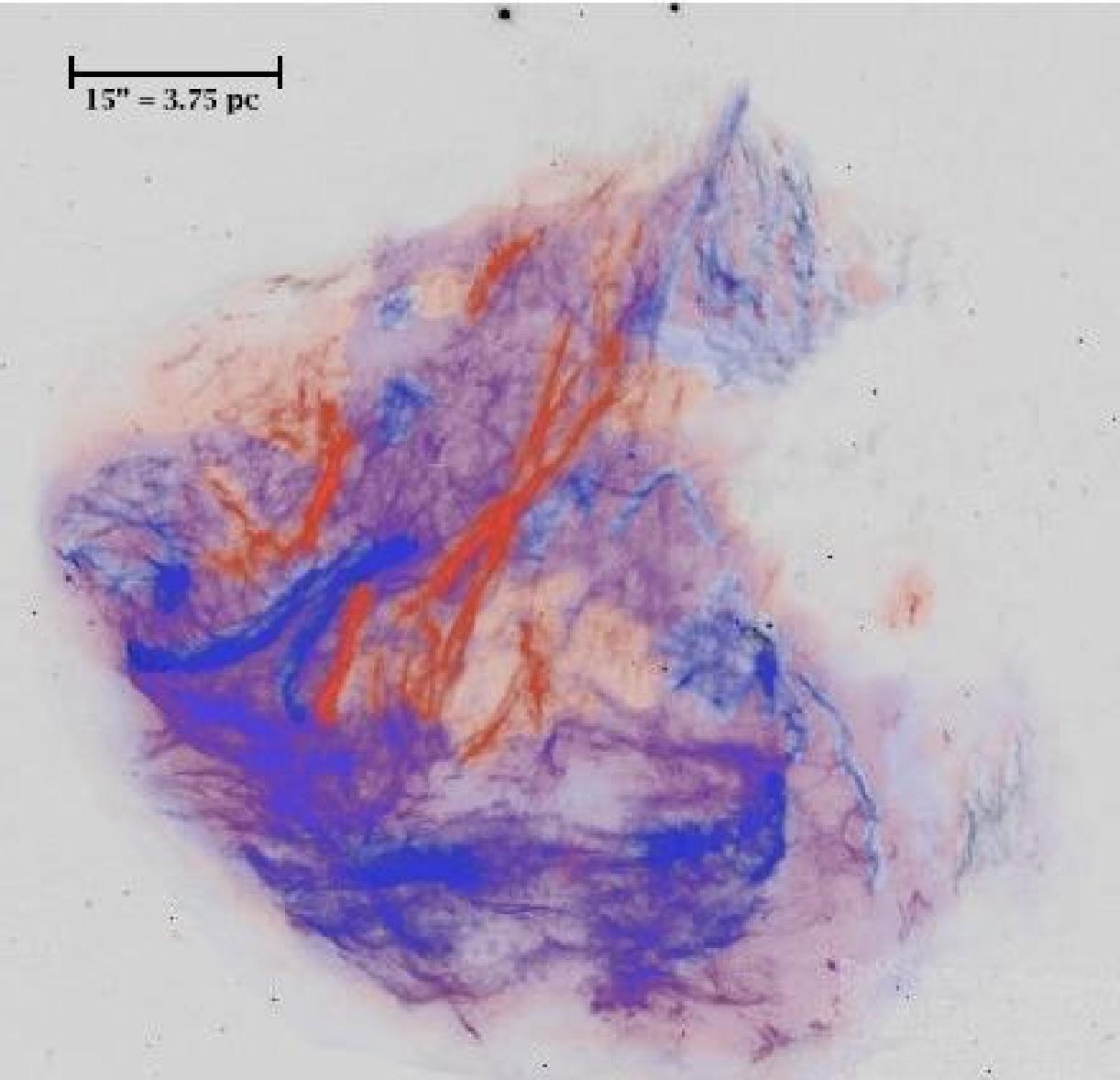}
\includegraphics[width=0.49\textwidth]{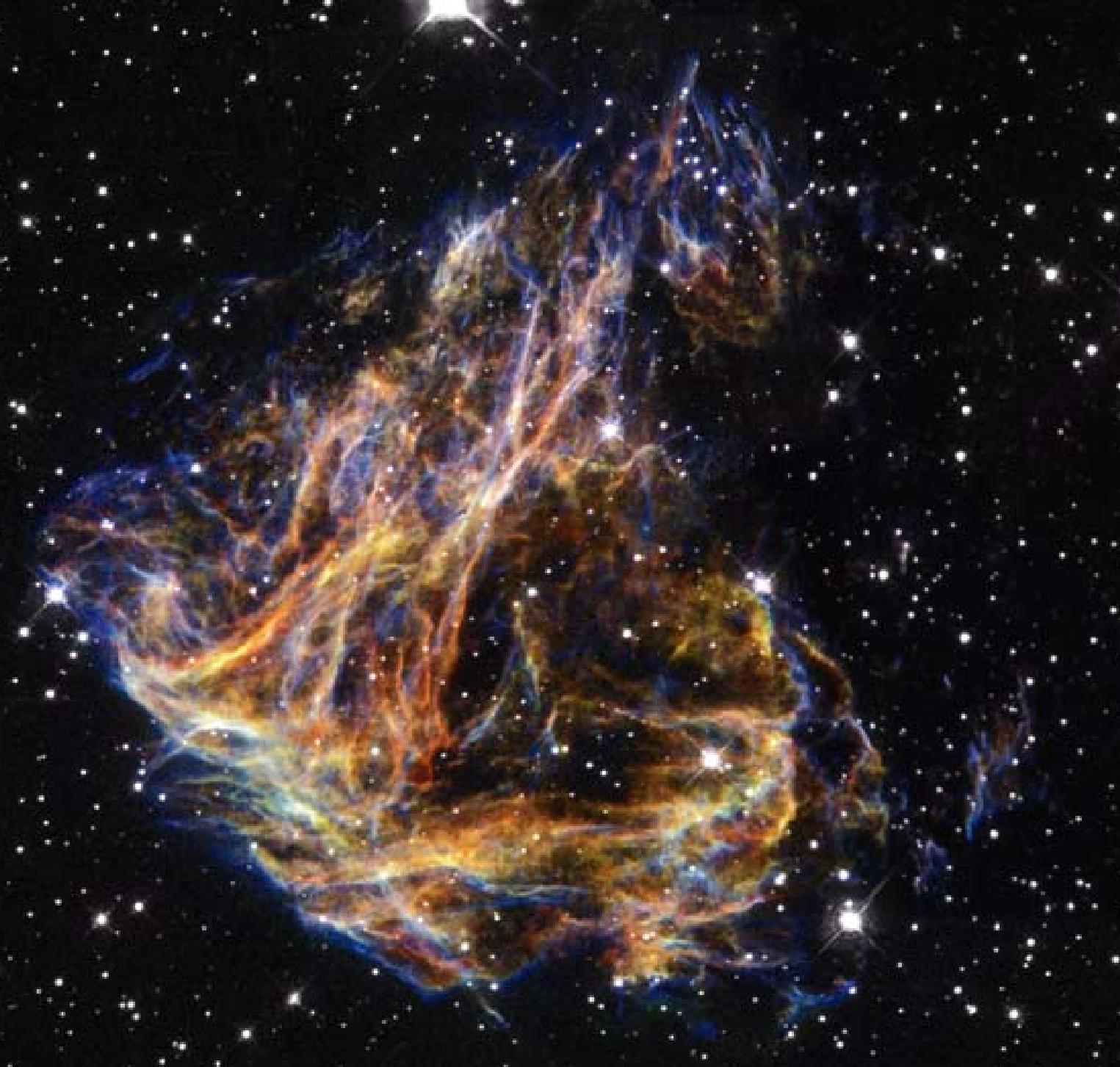}
\caption{(a) A 3D cartoon of N49 with blue and red filaments corresponding to the near and the far side of the remnant, respectively. (b) An overlay {\it HST} image of H$\alpha$ (red), [\sii] (yellow) and [\oiii] (blue) emission.} 
\label{fig:color}
\end{center}
\end{figure}

The enhanced [\oiii]/H$\alpha$ ratios at the edges of the SNR
appear to be at least as much due to a diminishment in the H$\alpha$ emission
as to an increase in [\oiii] emission. 
This morphology demonstrates an offset of [\oiii] emission toward the outer
edge of the remnant, seen also in the composite image of H$\alpha$, [\sii] 
and [\oiii] emission (Fig. \ref{fig:color}b).
We evaluate the offset by comparing the peaks of 
brightness profiles across the filaments in [\oiii] and H$\alpha$. 
Sample SB profiles demonstrating this offset are shown in 
Fig. \ref{fig:sbprofile}.
We find the offsets to range from 0 up to $\sim$ 0\farcs5, or 0.125 pc
(for a distance of 50 kpc; Feast 1999).

On small scales, we would expect an offset between the [\oiii] and
H$\alpha$ filaments because the emissivity and ionic concentration
of a given line depend on the temperature and thus the distance 
from the shock front.
In a plain, steady-shock model, the hydrogen recombination emission peaks 
roughly where [\oiii]
and [\oii] emission diminishes and [\oi] emission rises \citep{Cox72}.
The time scale needed for a post-shock region to evolve from 
[\oiii]-bright to [\oiii]-faint is $\sim$2000/$n_0$ yr, where $n_0$ 
is the pre-shock density in units of H-atom cm$^{-3}$ \citep{Raymond80}.
Therefore, we expect the post-shock [\oiii] emission peak and the 
H$\alpha$ emission peak to exhibit a displacement of 
$v \times (2000 {\rm yr}/n_0)$, where $v$ is the velocity
of the post-shock gas with respect to the shock front.
In case of an adiabatic (Sedov) shock, the post-shock gas expands downstream  
with $\onequarter$ of the shock velocity; thus the shocked gas in the clouds 
is observed to move with $v_c = \threequarters v_s$, where $v_s$ is the shock 
speed.

For a dense cloud with $n_0$ = 1000 cm$^{-3}$ (the lower limit for the filament
density calculated by \citet{Vancura92M}), 
with an observed expansion velocity $v_c$ = 250 km~s$^{-1}$, the displacement 
between [\oiii] and H$\alpha$
will be $1.7\times10^{-4}$ pc, or 0\farcs0007, which cannot be resolved 
by the {\it HST}.
For the shock with the same velocity driven into the diffuse regions of N49,
where the density is about $n_0$ of 10 cm$^{-3}$,
the offset between [\oiii] and H$\alpha$ can be raised to 0.017 pc,
or 0\farcs07. This value still cannot be resolved by the {\it HST}.

There are several other scenarios that would produce an offset between [\oiii]
and H$\alpha$ emission, each involving a departure from a steady-flow 
shock model.
Using observations from the MagIC camera at the Clay Telescope, 
\cite{Rakowski07} compared the locations of the [OIII] and H$\alpha$ peaks, 
finding separations of up to 0\farcs62. Our HST observations, which do not 
suffer 
from seeing distortions, show peak separations of up to 0\farcs5.  
\cite{Rakowski07} attributed these offsets to thermal instabilities dampened by
magnetic field, 
following the models of \cite{Innes92}, who associates the offsets with the  
presence of secondary shocks.  
These models have predicted maximum offsets of 3 $\times$ 10$^{17}$ cm,
which for the distance of the LMC yields roughly 0\farcs4 \citep{Rakowski07}. 
However, the physical conditions for the models ($v_s$=175 
km s$^{-1}$, $n_0$ = 1 cm$^{-3}$, $B_0$ = 3 $\mu$G) are different from the ones
in the LMC. 
As \citet{Rakowski07} point out, the abundances of the LMC will increase the 
cooling time by a factor of 3, which would tend to increase the offset by 
this factor. Our derived pre-shock density is 10 cm$^{-3}$, which is 10 times 
higher than the value used for the models, so the net result of 
these 2 effects is a decrease in the offset by a factor of 10/3, giving an 
offset of roughly 0\farcs15. 
\citet{Rakowski07} argued that the fast shocks seen in N49 made this the most 
probable scenario. However, without a substantial amount of magnetic support, 
the derived values are still smaller than the maximum observed offsets. 

An alternate scenario was presented by Raymond et al. (1980, 1983) for such
offsets in the Cygnus Loop. In this scenario, the offsets arise due to 
incomplete recombination zones behind a shock front driven into an 
inhomogeneous environment.
For a SNR in a cloudy medium, the slow shocks driven into dense 
clouds produce the [\oiii] and H$\alpha$ emission, while the fast 
shocks propagating into the less-dense intercloud medium produce 
X-ray emission.  The H$\alpha$ and [\oiii] lines are both emitted 
from shocked dense clouds, but their ``turn-on`` times are different
by $\sim$2000/$n_0$ yr.  As the fast shocks advance through the
intercloud medium and encounter clouds, [\oiii] emission will be
produced almost immediately in the freshly shocked clouds, and
the H$\alpha$ emission is produced by clouds that were shocked
$\sim$2000/$n_0$ yr earlier.  The displacement between the [\oiii]
and H$\alpha$ peaks is thus $\sim (v_{s,i} - v_{s,c}) \times 2000 
{\rm yr}/n_0$, where $v_{s,i}$ is the velocity of a fast shock
through the intercloud medium and $v_{s,c}$ is the velocity
of a slow shock through a dense cloud \citep{Fesen82}.

The velocity of the fast shock in the intercloud medium can be estimated
from the temperature of the post-shock, X-ray emitting gas, using
\begin{equation}
kT = \frac{3}{16} \mu m_{\rm H} v_{s,i}^{2} ,
\label{for:kT}
\end{equation}
where $m_{\rm H}$ is the mass of the hydrogen atom and $\mu$ is the mean 
molecular weight.
 For a fully ionized gas with He:H = 1:10, $\mu$ is 0.6.
 Adopting the higher temperature component $kT$ = 1.0 keV \citep{Park03}, we 
get a $v_{s,i}$ = 920 km s$^{-1}$.
The highest velocities seen in the dense shocked clouds, obtained from the 
echelle spectra (see \S 4.), are about 250 km s$^{-1}$. The post-shock gas in 
the clouds was accelerated to 0.75$v_{s,c}$.
The pre-shock density is 10 cm$^{-3}$. Therefore, the expected
displacement between the [\oiii] and H$\alpha$ peaks is 0.12 pc, or 0\farcs5.
This offset is in excellent agreement with the observation.

\section{Analysis of Filament Velocity Structure with Echelle Spectra}
To determine the exact slit positions of the echelle observations, we match 
the $SB$ profiles along the slit to those extracted from the {\it HST} 
H$\alpha$ image. 
The high angular resolution of the {\it HST} images allows us for 
the first time to see one-to-one correspondence between the kinematical and 
morphological features of N49 (Fig. \ref{fig:echelle}).

\subsection{Discrete Features}

\begin{figure}
\begin{center}
\includegraphics[width=0.85\textwidth]{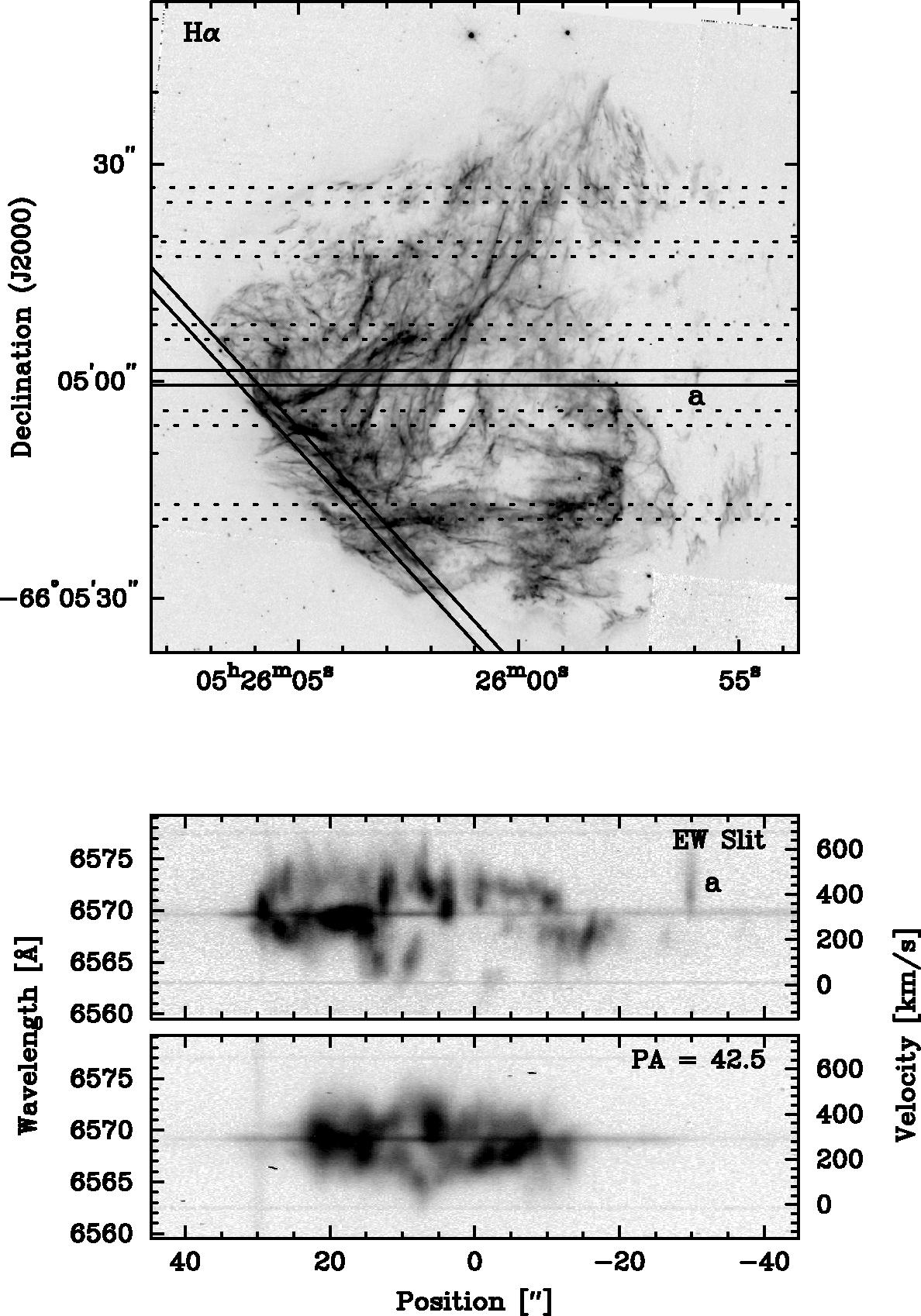}
\caption{Echellograms of N49 for E2000-1 and E1995-1 data, with corresponding 
positions on the H$\alpha$ image marked by the solid horizontal and tilted 
lines, respectively.The feature a in the E2000-1 echelle spectrum and the 
optical image is a high-velocity clump discussed in \S 4.1. The dashed lines 
mark the positions of E1986-1, -2,-3, -4 and -5 (bottom to top) spectra taken 
by \cite{CK88}.} 
\label{fig:echelle}
\end{center}
\end{figure}

The echellograms in Figure \ref{fig:echelle} show 
discrete broad emission features as well as a narrow component extending continuously along the slit.
The narrow component, detected by \citet{CK88} and \citet{Vancura92M} shows 
$SB$ enhancements within the boundary of the SNR, indicating physical association; we therefore adopt its heliocentric velocity ($v_{\rm hel}$) of 315 km s$^{-1}$ as the SNR's systemic velocity. 
Due to our extended slit coverage, we are able to pin down the location of this enhanced narrow component in the SNR. The $SB$ variation of this narrow component is similar to that of the diffuse H$\alpha$ emission, brighter in the east and fainter in the west.
 We therefore associate this narrow component with the faint diffuse H$\alpha$ component identified morphologically in \S 3.1. 
The narrow velocity width and the pre-shock location of this narrow compnent 
are consistent with its origin as recombination radiation of hydrogen ionized 
by the UV precursor of the SNR shocks.


We also observe broad emission features which indicate the bulk expansion of about 250 km s$^{-1}$.
Some of the broad emission features have a ``head-tail'' structure; the bright emission at a head has a smaller radial
velocity offset ($\Delta v$) from the systemic velocity and the fainter emission in the tail stretches to larger $\Delta v$. Among the discrete broad emission features, the brighter ones have smaller $\Delta v$, 
ranging from 0 to 160 km s$^{-1}$ at the ``head'' and up to about 
250 km s$^{-1}$ at the ``tail''.
They can be well matched to individual bright filaments.

The structure of these emission features can be explained by the interaction between the shock and the encountered dense gas.
The shock gets dampened more by the dense gas in the filaments than in the lower-density cloudlets. 
Therefore, the post-shock gas in the inner part of the dense bright regions was propelled to velocities significantly lower than the original shock velocity, thus producing smaller velocity spreads.
The presence of gas at systemic velocity suggests that the gas in these regions retarded the shock sufficiently to prevent its propagation throughout the whole filament, leaving the innermost gas unshocked.
Because it emits the H$\alpha$ recombination radiation, it must have been ionized by the UV precursor.
The fainter ``tail'' emission comes from the outer layers of these dense filaments that encounetered the undampened shock and have been accelerated to higher velocities.


The less dense gas, which can be identified on echelle spectra as fainter emission component, can be accelerated by the schock to very high velocities. The lower-density gas is unable to significantly slow down the shock, and the post-shock gas is consequently accelerated to higher speeds. We observe velocity spreads of up to 300 km s$^{-1}$. Some of these cloudlets have no systemic velocity components, which implies that all of the gas is shocked.

A few cloudlets have velocity spreads that exceed 300 km s$^{-1}$. 
One example is a faint emission knot seen at the edge of the remnant in observation E2000-1, which has been marked in Figure \ref{fig:echelle} as feature a.
The clump has a radial velocity spread of 350 km s$^{-1}$.
Due to its location at the edge of the remnant, it likely has a significant tangential velocity component and thus the actual velocity of the blob is even higher. Assuming the two components to be comparable, the actual velocity of some of the gas in this blob could be up to $\sim$ 500 km s$^{-1}$.


\subsection{Overall Kinematic Structure} 

The observed distinct broad emission features have implications for the physical structure of the SNR.
The echelle spectra do not suggest the presence of an expanding shell, which would demonstrate itself as a simple bow-shaped line image.
 The absence of this expanding shell signature in the echelle spectra implies that the observed filaments are not projection effects of an expanding sheet of gas, but actual long ribbon-like structures which indicate clumpiness in the surrounding gas prior to the explosion. 

We have taken echelle spectra in the southern part of the remnant (E2000-2, E2000-3) to see whether the arc-like filament in the south is Balmer line dominated. We can however observe [\nii] $\lambda$6584 line in the spectra, and thus conclude that the arc is not Balmer dominated.

In observations E1986-2 and E2000-1, we observe very few broad component features at systemic velocity around the center of explosion determined from X-ray and radio data. In the blueshifted part of the echelle spectra, we see no broad features at all in E1986-2, and only few faint ones in E2000-1. In the redshifted part, we observe a number of broad features in both spectra, but only one of these in E2000-1 has the lower end of its velocity spread at the systemic velocity. The rest of the broad features are seen at higher velocities. Such spectra suggest the presence of a central cavity; its position matches the X-ray and radio observations. The asymmetry of the emission in this central region implies uneven density distribution of matter prior to the explosion, with higher density at the far side of the remnant.

The comparison of the echelle spectra with optical morphology of the SNR allows us to see three-dimensional geometry of the remnant. Using the echelle spectra, we were able to discern whether individual filaments are on the approaching or the receding side of the remnant. Assigning blue and red colors to the blue- and redshifted filaments, respectively, we have constructed a 3D image of N49 (Fig. \ref{fig:color}a).  
For the most part, we see the near and the far side of the remnant superimposed, but we observe a number of prominent red and blue filaments. The most prominent redshifted filaments are the straight long filaments extending form the SE Ridge northwards. The majority of the outer filaments is blueshifted, for example a filament extending westward from the SE Ridge. 

\section{Comparison of X-ray and Optical Morphology}
To better trace the relationship of the warm ionized gas in the
optically-emitting filaments with the hot shocked gas, we compared
the {\hst} images with data from {\it XMM-Newton} and {\it Chandra}.
All of the X-ray images show a bright, point-like X-ray source to the
north corresponding to the location of the soft gamma-ray repeater
SGR 0525-66, as noted by \cite{Rotschild94}. This bright source appears
between optically bright features, in a relatively void optical
region, and it is not associated with any optical source.

\begin{figure}
\begin{center}
\includegraphics[width=0.49\textwidth]{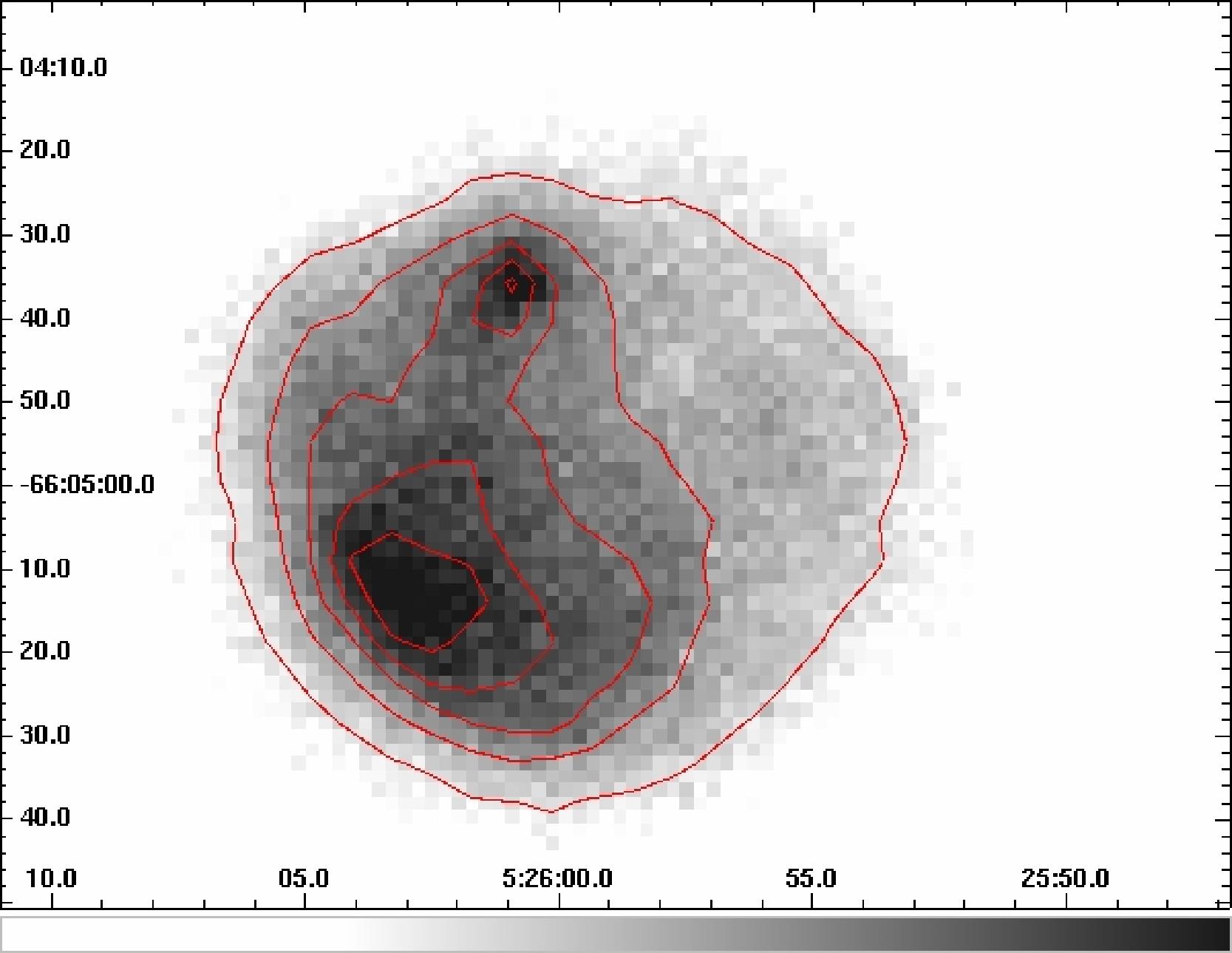}
\includegraphics[width=0.49\textwidth]{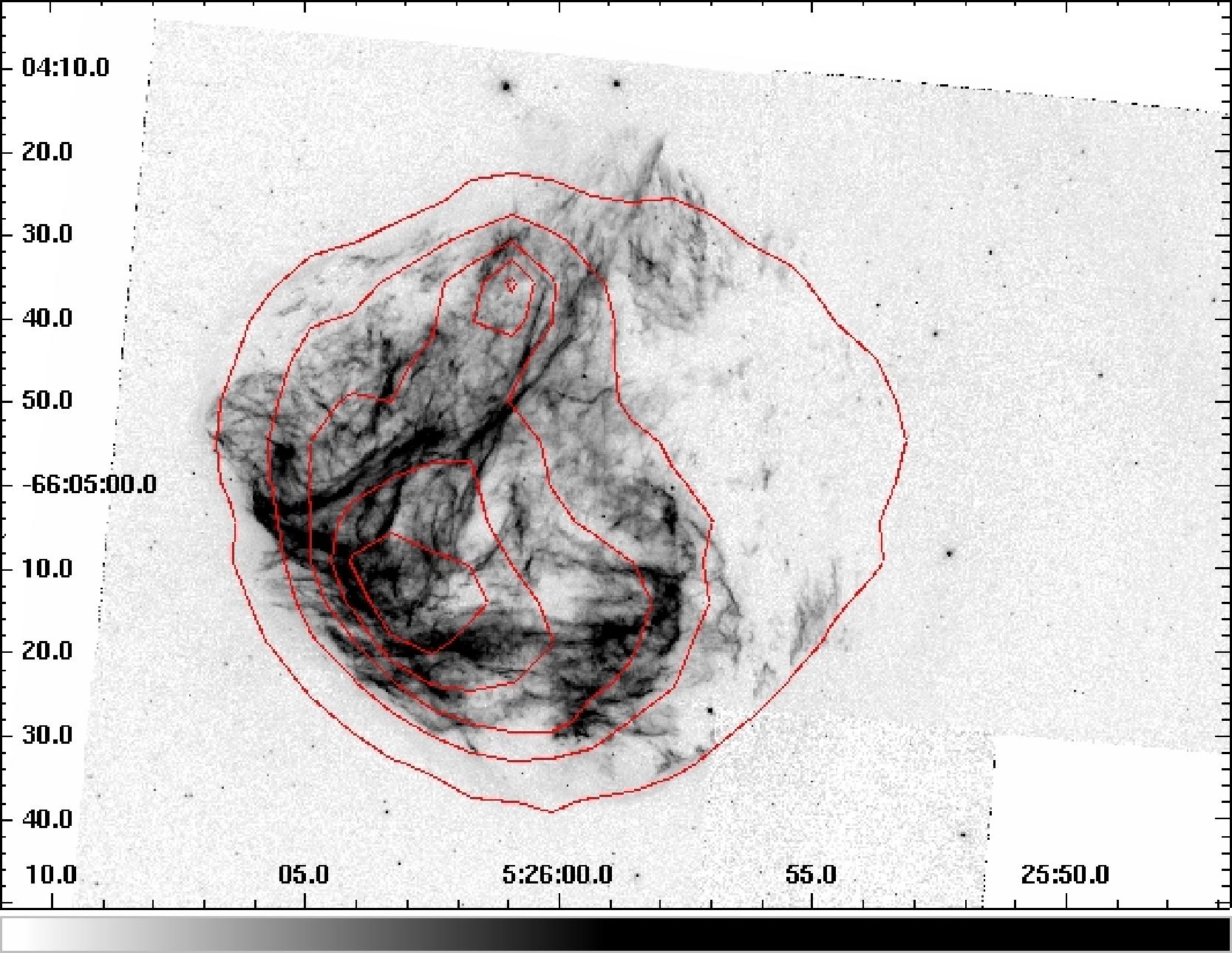}
\caption{(a) Smoothed contours of {\it XMM-Newton} X-ray data; (b) The same 
{\it XMM-Newton} contours overlaid on the optical H$\alpha$ image from 
{\it HST}.  This image was binned in pixels of 1$\farcs$6 $\times$ 1$\farcs$6. 
Contours plotted on the image range from 3$\sigma$ -- 30$\sigma$ 
above the background in 5 evenly spaced linearly increasing contours.} 
\label{fig:xmm}
\end{center}
\end{figure}

The X-ray image of N49 from {\it XMM-Newton} is presented in Figure
\ref{fig:xmm}a.  The corresponding X-ray contours are overlaid on 
the H$\alpha$ observation from {\hst} in Figure \ref{fig:xmm}b.  
The 2$\sigma$ contour
provides a nicely defined boundary to the expanding shell. In general,
the comparison of {\it XMM-Newton} and optical images reveals a strong
relationship between X-ray brightness and the regions of highest optical
flux. As with the optical brightness, the X-ray flux declines toward the
west, although the western side of the SNR is still detectable. To the
southeast, the steeper X-ray contours indicate a substantial X-ray
brightening corresponding to the bright SE ridge in optical; this
emission appears to be resolved without any obvious point source. Westward from
this X-ray peak, the central cavity described in the velocity structure
analysis (\S4) appears somewhat brighter in X-rays than do regions
to the west and north of the SNR.

\begin{figure}
\begin{center}
\includegraphics[width=0.49\textwidth]{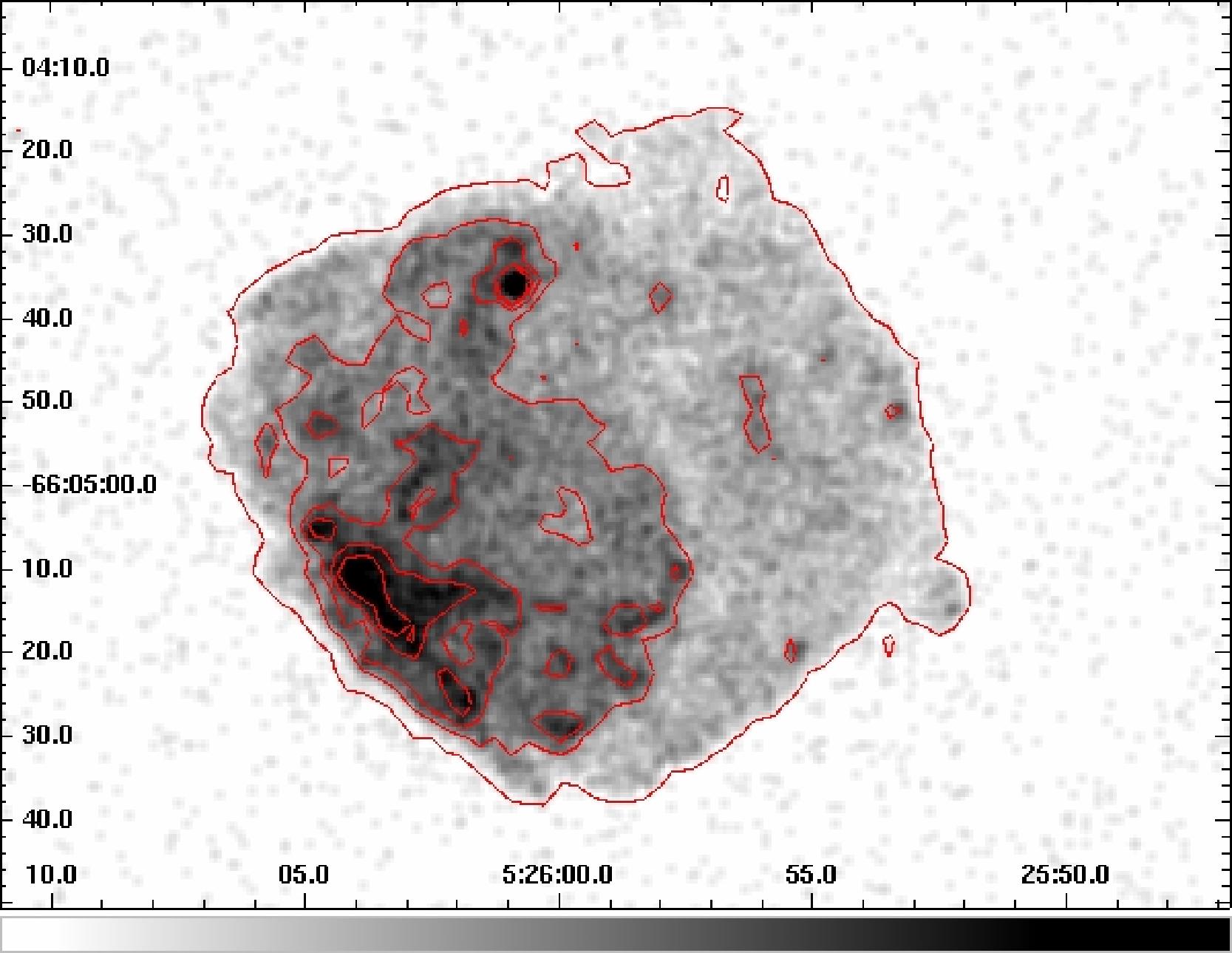}
\includegraphics[width=0.49\textwidth]{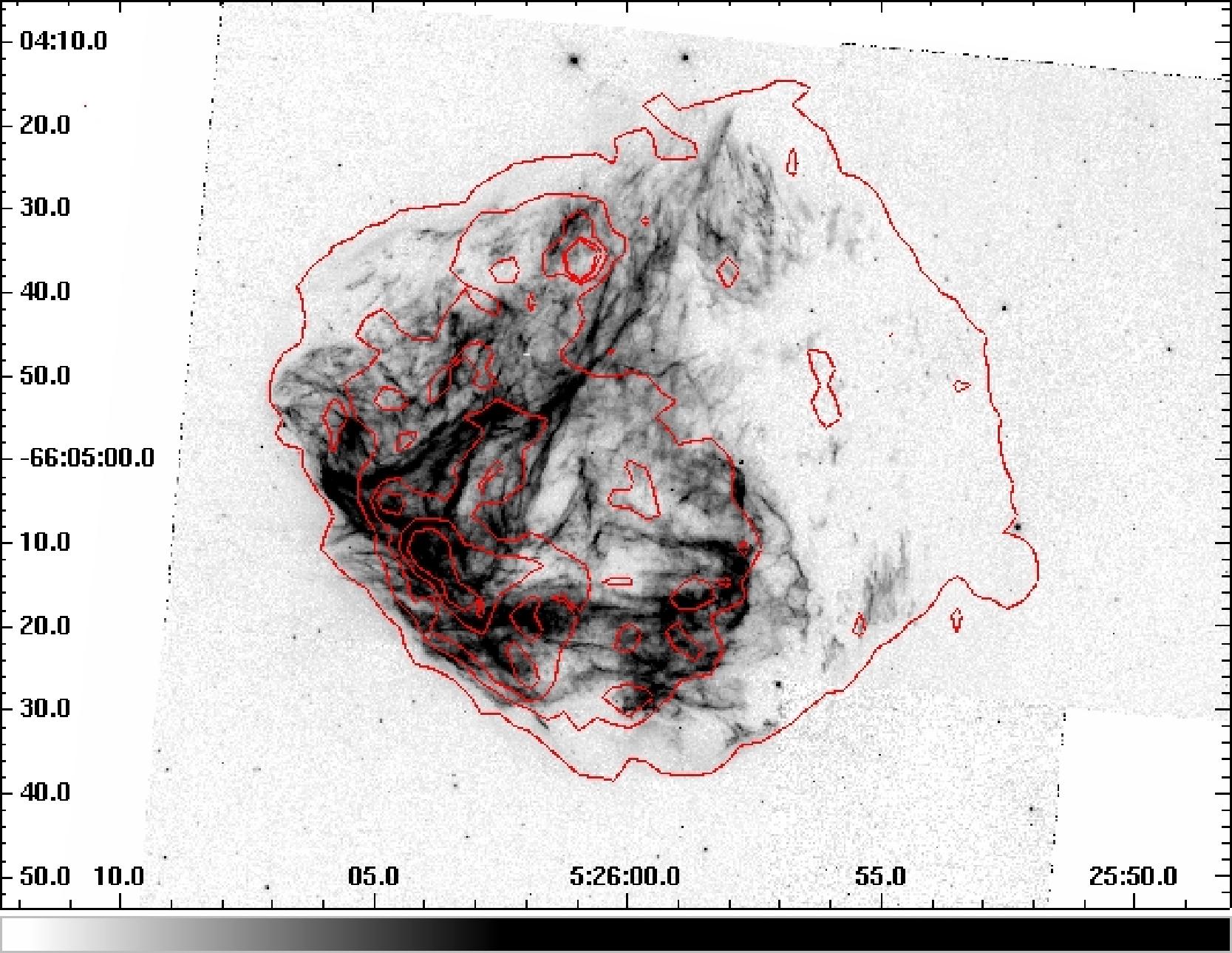}
\caption{(a) Smoothed contours of {\it Chandra} ACIS X-ray data; (b) The same 
contours overlaid on the optical H$\alpha$ image from {\it HST}. 
This image was  processed with a 2 pixel gaussian smoothing. Contours plotted 
on the image range from 2$\sigma$ -- 100$\sigma$ above the background in 5 
evenly spaced linearly increasing contours.} 
\label{fig:chandra}
\end{center}
\end{figure}

A recent short {\it Chandra} exposure enables an X-ray/optical comparison
at higher resolution.  The X-ray image of N49 from {\it Chandra} is
presented in Figure \ref{fig:chandra}a.  Again, the corresponding X-ray 
contours are overlaid on the H$\alpha$ observation from {\hst} in 
Figure \ref{fig:chandra}b for a
multiwavelength comparison. As seen for the {\it XMM-Newton} observation,
the overall X-ray surface brightness distribution is similar to that in
H$\alpha$, in agreement  with earlier {\it Chandra} ACIS observations by
\citet{Park03}. In particular, the ACIS observations show X-ray features
which roughly correspond to the brightest groups of optical filaments,
such as the SE ridge and the filaments that project from this region to
the northwest and southwest. Similarly, the ``patches" of optical
emission on the western face of the remnant are accompanied by a minor
X-ray enhancement in that region. The X-ray emission also outlines the
optical cavity region, with the emission decreasing toward the interior
of the region.  Brighter X-ray emission traces the boundary with a radius
of curvature similar to that expected from both the optical and echelle
data.   On the eastern and southern sides of the remnant, the faint
X-ray extent matches very well the SNR boundary seen in diffuse
H$\alpha$ emission.

In addition to this morphological study, we examined the X-ray spectra
using {\it XMM-Newton} data for three regions corresponding to significant
optical features.  These regions include one covering the SE Ridge
of optical emission, one along the optically faint region in the northwest,
and a third region corresponding to the cavity described in \S4.  For each 
region, spectra were extracted for each of the three \xmm\ instruments 
(MOS1, MOS2, and pn) and fit jointly with spectral models for a thermal 
plasma modified by a model of photoelectric absorption.

The primary plasma model used is a plane-parallel shock, non-equilibrium 
ionization model ({\it vpshock} in XSPEC). This model presumes a simple 
shock structure with a constant postshock electron temperature $kT$.  
Ionization timescales are modeled as a broad distribution described by 
the ionization age $\tau=n_e t_s$, where $n_e$ is the post-shock electron 
number density and $t_s$ is the time since the plasma was shocked. This 
model provides a reasonable, if simplified, approximation to shocked 
plasma conditions for relatively high-temperature shocks in limited 
spatial regions of an SNR \citep{BLR01}.  

When describing broad spatial regions of an SNR, as we do here, one should 
keep in mind that the spectra arise from a range of temperatures and densities
within that region. A single-temperature approximation is therefore of only
limited utility in correctly describing the ``characteristic" physical 
conditions.  In many cases, it is necessary to use at least two linked models
at different temperatures in order to better fit a spectrum which is in 
reality produced by a superposition of shocks in different conditions. 
However, particularly in smaller regions, too few counts are often present
to constrain a large set of parameters, as for instance those created by 
adding extra temperature components. The model fits are therefore a compromise:
we use as few parameters as possible to characterize a given dataset, adding 
complexity as the number of counts increases sufficiently to require - and
support - additional spectral components.

The best-fit spectral models for the SNR and for the selected spatial regions
are given in Table~4.  Listed in this table are the plasma model components 
used and the fitted values for absorption column density ($N_H$), post-shock 
temperature ($kT$), abundances of significant elements as a fraction of solar 
values ($O/O_{\sun}$, etc.); and ionization age ($\tau$).  Also listed are 
normalization values, proportional to the emission measure, for each of the 
\xmm\ instruments ($A_{MOS1}$, etc).  Finaly, for each model fit the statistical 
goodness-of-fit values are given: reduced $\chi^2$ ($\chi^2_{\rm red}$) and 
degrees of freedom (dof).  Considering the associated errors and the lower 
spatial resolution of {\it XMM-Newton}, the results agree well with fits to
{\it Chandra} ACIS data made by \citet{Park03}.

The X-ray spectra reflect the morphological differences between the
eastern and western sides of the SNR.  To the west, a single-component
plasma model can adequately describe the spectrum. Likewise, the cavity 
region requires only a single plasma component for a reasonable fit. To 
the southeast, on the other hand, the spectrum requires two plasma 
components for a reasonable fit ($\chi^2$/dof=1122/743). Notably, a 
combination of plasma and power-law components provides a considerably 
poorer fit ($\chi^2$/dof=1489/744), allowing us to rule out, at the 90\%
confidence level, the possibility that the high-energy component of
the spectrum is due to a nonthermal contribution.

\citet{Park03} interpret this two-component fit to emission from
the southeastern SNR as follows: the low-temperature component
arising from the forward shock moving slowly into dense clouds, and
the high-temperature component originating from the reheating of
postshocked material by shocks reflected from dense clumps.  Based
on our velocity analysis, however, we suggest a slight modification
of this picture.  Rather than having two distinct temperature regimes,
this region of the SNR, we suggest, is the location of a distribution
of shock velocities.  This picture is supported by the broad range of
measured optical expansion velocities, as well as by the modelling of
optical line emission by \citet{V92}.  The two-temperature fit to
the X-ray emission, therefore, can be regarded as an approximation
to emission at a wide range of temperatures, generated by forward
and reflected shocks at various speeds.  These variations in local
shock speed presumably arise from the encounter of the overall blast
wave with material at different densities, as would be expected where
the SNR is encountering the molecular cloud  \citep{B97}.

\section{Summary}

We have conducted a multiwavelength study of SNR N49 using {\it HST} images 
in H$\alpha$, [\oiii] and [\sii] emission lines, archival {\it Chandra} 
and {\it XMM-Newton} images, and compared the morphological features to 
SNR kinematics, using the high-dispersion echelle spectra obtained at CTIO. 

All emission line images reveal a highly filamentary morphology, with most
prominent features being the SE Ridge, long filaments extending NW and a 
network of filaments in the SE. The optical emission from the western part 
of the remnant is significantly lower.
We also detect a diffuse H$\alpha$ emission component protruding beyond the 
shock boundary in all regions except for the west, where this emission is 
absent or undetected. We identify this diffuse component as the recombination 
radiation from the gas ionized by the UV precursor.
Using H$\alpha$ surface brightness, we have estimated the rms electron 
densities in different regions of the remnant. 

Good correlation between [\sii] and H$\alpha$ morphology is demonstrated 
in the uniformity of the ratio map, with the mean value of 0.75 $\pm$ 0.05, 
and overall range of 0.4-1.4.
[\oiii] images and a wide range of values on the [\oiii]/H$\alpha$ ratio map 
manifest morphological differences between the [\oiii] and H$\alpha$ emission. 
The most prominent feature, an offset between [\oiii] and H$\alpha$ bright 
filaments, is demonstrated on the ratio map as an enhancement of 
[\oiii]/H$\alpha$ ratio toward the outer edge of the image, where ratios 
reach between 0.8 and 1.5, with a few patches reaching up to 2.5. We observed 
the highest offset of 0\farcs5.
We propose that such an offset would arise in one of the two scenarios:
(a) The offset is due to the thermal instabilities cushioned by magnetic fields
and the associated secondary shocks. The fast shocks observed in N49 make this
a likely scenario; however, a substantial amount of magnetic support is needed
to account for the values of maximum observed offsets. 
(b) The offsets arise due to shocks propagating into a clumpy medium. The 
fast shocks just encountering the cloud produce the [\oiii] emission, while 
H$\alpha$ comes from previously shocked clouds that have cooled sufficiently. 
The offset produced by this model matches the observations.

We have also carried out a detailed study of the remnant kinematics using 
high-dispersion echelle spectra at 7 different slit positions throughout the 
SNR. The spectra reveal a clumpy velocity structure, with discrete broad
emission features matching well the positions of filaments across the 
slit, with radial velocity spreads reaching up to 160 km s$^{-1}$. We also 
detect a number of fainter clumps with very high radial velocity spreads that 
reach 350 km s$^{-1}$, and estimate the actual velocity of these clumps to 
be up to 500 km s$^{-1}$.
In addition, we detect a narrow component extending throughout the slit, 
and we associate it with the diffuse H$\alpha$ component identified 
morphologically.

We do not observe a bow-shaped signature of an expanding shell in the spectra. 
This indicates a true filamentary characteristic of the remnant and implies a
clumpy structure of the gas surrounding the star prior to the explosion.

We also confirm the presence of the cavity in the remnant which matches the 
X-ray and radio data, and observe the asymmetric expansion of the gas in this 
region, which we associate with irregular density distribution of the 
surrounding matter prior to the explosion. 
Matching discrete emission features to actual filaments allows us to identify 
the blue- and redshifted filaments. Through a detailed comparison between the 
echelle spectra and the optical images, we have constructed a rough 3D map of 
the remnant.

We have compared the archival {\it Chandra} and {\it XMM-Newton} images to the 
H$\alpha$ morphology. The X-ray morphology of N49 correlates well with the 
general trends seen in optical wavelengths. Even though we do not observe 
filamentary structure in X-rays, the brightest emission in X-rays comes from 
the SE, and the emission diminishes towards the west. The X-ray boundary 
matches the extent of optical filaments. Unlike in the optical, we observe the 
whole spherical envelope of the SNR in X-rays.

In addition to the images, we have carried out spectral analysis of the 
remnant. While the cavity region as well as the western side of the remnant 
can each be fit by a single plasma component, the eastern and SE regions 
require two component plasma. We rule out non-thermal radiation with 90$\%$ 
confidence.
We suggest the 2 plasma component to be an approximation to a range of 
temperatures generated by forward and reflected shocks at various speeds.





\acknowledgements

The authors gratefully acknowledge the support of the Dean of the College 
of Liberal Arts and Sciences of the University of Illinois for YHC.  
RMW and YHC acknowledge support from STScI grant STI GO-8110. 
This material is based upon work supported by the National 
Aeronautics and Space Administration under grant NNG05GC97G 
issued through the Long-Term Space Astrophysics Program. We would like
to thank the referee for useful suggestions to improve the paper.



Facilities: \facility{HST (WFPC2)}, \facility{CXO (ACIS)},
\facility{XMM (EPIC)}, \facility{Blanco}.





\clearpage
\begin{deluxetable}{lccccc}
\tablecaption{Optical {\it HST} WFPC2 Observations}
\tablehead{
\colhead{Observation} &
\colhead{Date} &
\colhead{Line} &
\colhead{Filter}&
\colhead{Exposure} &
}
\startdata

1 & 2000 Jul 13 & \ha\ $\lambda$6563 & F656N & 2 $\times$ 500 s  \\
2 & 2000 Jul 14 & [\sii] $\lambda\lambda$6716,6731 & F673N & 2 $\times$ 600 s \\
3 & 2000 Jul 14 & [\oiii] $\lambda$5007 & F502N & 720+637+800 s \\

\enddata
\label{ObsHST}
\end{deluxetable}

\begin{deluxetable}{lccccc}
\tablecaption{CTIO 4m Echelle Observations}
\tablehead{
\colhead{Observation} &
\colhead{Date} &
\colhead{Exposure} &
\colhead{Location} &
\colhead{Position Angle}
}
\startdata


E1986-1 & 1986 Nov 20 & 300 s & N49 pos 1 & 90 \\
E1986-2 & 1986 Nov 20 & 300 s & N49 pos 2 & 90 \\
E1986-3 & 1986 Nov 20 & 300 s & N49 pos 3 & 90 \\
E1986-4 & 1986 Nov 20 & 300 s & N49 pos 4 & 90 \\
E1986-5 & 1986 Nov 20 & 300 s & N49 pos 5 & 90 \\

E1995-1 & 1995 Jan 19 & 2$\times$600 s & N49 IRS pos & 42.5 \\

E2000-1 & 2000 Dec 10 & 600 s & N49 & 90 \\
E2000-2 & 2000 Dec 10 & 300 s & N49 south & 90 \\
E2000-3 & 2000 Dec 10 & 300 s & N49 south, 3$''$N & 90 \\

\enddata
\label{Observations}
\end{deluxetable}

\begin{deluxetable}{lcccc}
\tablecaption{X-ray fluxes for the regions normalized by area}
\tablehead{
\colhead{ } &
\colhead{Area} &
\colhead{Avg. Flux} &
\colhead{Norm. Flux} & \\
\colhead{} &
\colhead{arcmin$^2$} &
\colhead{ergs cm$^2$ s$^{-1}$} &
\colhead{ergs cm$^2$ s$^{-1}$ arcmin$^2$} &
}
\startdata

SE Limb & .269 & 2.23 $\times 10^{-11}$ & 8.3 $\times 10^{-11}$ \\
NW Limb & .300 & 9.47 $\times 10^{-12}$ & 3.2 $\times 10^{-11}$ \\
Cavity & .097 & 9.54 $\times 10^{-12}$ & 9.8 $\times 10^{-11}$ \\

\enddata
\label{NormFluxes}
\end{deluxetable}

\begin{deluxetable}{l|ccccccccc}
\tabletypesize{\scriptsize}
\tablecaption{Best-fit spectral models to XMM data for N49 and sub-regions}
\tablehead{
\colhead{Parameter} &
\multicolumn{2}{c}{Whole SNR} &
\multicolumn{2}{c}{SE Limb} &
\colhead{NW Limb} &
\colhead{Cavity}
}
\startdata
component  & \multicolumn{2}{c}{vpshock + vpshock} & \multicolumn{2}{c}{vpshock
+ vpshock} & vpshock & vpshock  \\[6pt]
$N_{\rm H}$ (cm$^{-2}$) & \multicolumn{2}{c}{2.9$^{+0.1}_{-0.1} \times10^{21}$}
        & \multicolumn{2}{c}{3.2$^{+0.3}_{-0.3} \times10^{21}$}
        & 2.3$^{+0.1}_{-0.1} \times10^{21}$  & 2.2$^{+0.1}_{-0.1} \times10^{21}$
\\[6pt]
$kT$ (keV) & 0.41$^{+0.01}_{-0.01}$ & 0.98$^{+0.01}_{-0.01}$
        &  0.40$^{+0.03}_{-0.03}$ & 1.02$^{+0.01}_{-0.03}$
        &  0.66$^{+0.01}_{-0.02}$  & 0.54$^{+0.01}_{-0.01}$ \\[6pt]
 O/O$_\sun$  & \multicolumn{2}{c}{0.24$^{+0.01}_{-0.01}$}
        & \multicolumn{2}{c}{0.21$^{+0.01}_{-0.01}$}
        &  0.65$^{+0.05}_{-0.05}$ & 0.38$^{+0.03}_{-0.03}$  \\[6pt]
 Ne/Ne$_\sun$ & \multicolumn{2}{c}{0.31$^{+0.01}_{-0.01}$}
        & \multicolumn{2}{c}{0.29$^{+0.01}_{-0.02}$}
        & 0.62$^{+0.06}_{-0.05}$  &  0.38$^{+0.04}_{-0.02}$ \\[6pt]
 Mg/Mg$_\sun$ & \multicolumn{2}{c}{0.31$^{+0.01}_{-0.02}$}
        & \multicolumn{2}{c}{0.25$^{+0.02}_{-0.03}$}
        & 0.30$^{+0.06}_{-0.01}$  &  0.23$^{+0.03}_{-0.03}$ \\[6pt]
 Si/Si$_\sun$ & \multicolumn{2}{c}{0.57$^{+0.02}_{-0.02}$}
        & \multicolumn{2}{c}{0.65$^{+0.05}_{-0.04}$}
        & 0.41$^{+0.05}_{-0.06}$  &  0.48$^{+0.08}_{-0.04}$ \\[6pt]
 S/S$_\sun$ & \multicolumn{2}{c}{ 0.67$^{+0.05}_{-0.05}$}
        & \multicolumn{2}{c}{ 0.9$^{+0.1}_{-0.1}$}
        & 0.7$^{+0.2}_{-0.2}$  &  1.1$^{+0.5}_{-0.1}$ \\[6pt]
 Fe/Fe$_\sun$ & \multicolumn{2}{c}{0.34$^{+0.01}_{-0.01}$}
        & \multicolumn{2}{c}{0.34$^{+0.01}_{-0.02}$}
        & 0.24$^{+0.01}_{-0.02}$  & 0.16$^{+0.01}_{-0.02}$  \\[6pt]
$\tau$ (cm$^{-3}$ s) & 3.1$^{+0.1}_{-0.3} \times10^{11}$ & $>1.0 \times10^{13}$
        & 2.8$^{+0.4}_{-0.2} \times10^{11}$ & $>1.2 \times10^{13}$
        & 9$^{+1}_{-1} \times10^{11}$  &  1.2$^{+0.2}_{-0.1} \times10^{12}$ \\[6pt]
$A_{\rm MOS1}$ (cm$^{-5}$) & 4.8$^{+0.1}_{-0.1} \times10^{-2}$
        & 1.6$^{+0.1}_{-0.1} \times10^{-2}$
        & 2.8$^{+0.5}_{-0.4} \times10^{-2}$ &  6.2$^{+0.3}_{-0.2} \times10^{-3}$
        & 5.9$^{+0.1}_{-0.2} \times10^{-3}$  & 9.3$^{+0.4}_{-0.3} \times10^{-3}$
\\[6pt]
$A_{\rm MOS2}$ (cm$^{-5}$) &  5.6$^{+0.1}_{-0.1} \times10^{-2}$
        &  1.6$^{+0.1}_{-0.1} \times10^{-2}$
        & 3.0$^{+0.5}_{-0.4} \times10^{-2}$ &  6.0$^{+0.3}_{-0.3} \times10^{-3}$
        & 6.3$^{+0.2}_{-0.2} \times10^{-3}$   & 1.0$^{+0.1}_{-0.1} \times10^{-2}$
\\[6pt]
$A_{\rm pn}$ (cm$^{-5}$) &  5.8$^{+0.1}_{-0.1} \times10^{-2}$
        &  1.6$^{+0.1}_{-0.1} \times10^{-2}$
        & 3.1$^{+0.4}_{-0.3} \times10^{-2}$ &  6.8$^{+0.2}_{-0.3} \times10^{-3}$
        & 6.3$^{+0.3}_{-0.1} \times10^{-3}$  &  1.0$^{+0.1}_{-0.1} \times10^{-2}$
\\[6pt]
$\chi^2_{\rm red}$ & \multicolumn{2}{c}{2.13} & \multicolumn{2}{c}{1.54}  & 1.32
 & 1.56    \\
dof & \multicolumn{2}{c}{875}  & \multicolumn{2}{c}{625}  & 485  & 435  \\
\enddata
\tablecomments{Spectra cover the range between 0.3-8.0 keV. A is the
normalization constant, defined as
$A = 10^{-14}\int n_{\rm e}n_{\rm H}dV/(4 \pi D^2)$, where $n_{\rm e}$ and 
$n_{\rm H}$ are electron and proton densities, respectively, and $D$ is the 
distance, all in cgs units. Elements not listed are fixed to the LMC mean 
abundance. Errors give the
90\% confidence range for each fit parameter. Errors in $N_{\rm H}$, $kT$,
$\tau$ and $A$ are estimated with all abundances except Fe fixed at their
best-fit values; abundance errors are estimated with $N_{\rm H}$, $kT$ and
$\tau$ fixed.}
\label{tab:snrxspec}
\end{deluxetable}

\clearpage


\end{document}